\documentclass[sigchi]{acmart}
\AtBeginDocument{%
  \providecommand\BibTeX{{%
    \normalfont B\kern-0.5em{\scshape i\kern-0.25em b}\kern-0.8em\TeX}}}

\setcopyright{acmcopyright}
\copyrightyear{2022}
\acmYear{2022}
\acmDOI{10.1145/3544794.3558462}

\acmConference[ISWC '22]{The 2022 International Symposium on Wearable Computers}{September 11--15, 2022}{Cambridge, United Kingdom}
\acmBooktitle{The 2022 International Symposium on Wearable Computers (ISWC '22), September 11--15, 2022, Cambridge, United Kingdom} 
\acmPrice{15.00}
\acmISBN{978-1-4503-9424-6/22/09}

\usepackage{multirow}
\usepackage[shortlabels]{enumitem}

\begin{document}

\title{Non-contact, real-time eye blink detection with capacitive sensing}


\author{Mengxi Liu}
\email{mengxi.liu@dfki.de}
\affiliation{%
  \institution{German Research Center for Artificial Intelligence}
  \streetaddress{Trippstadter Str. 122}
  \city{Kaiserslautern}
%   \state{Germany}
  \country{Germany}
  \postcode{67663}
}

\author{Sizhen Bian}
\email{sizhen.bian@dfki.de}
\affiliation{%
  \institution{German Research Center for Artificial Intelligence}
  \streetaddress{Trippstadter Str. 122}
  \city{Kaiserslautern}
%   \state{Germany}
  \country{Germany}
  \postcode{67663}
}

\author{Paul Lukowicz}
\email{paul.lukowicz@dfki.de}
\affiliation{%
  \institution{German Research Center for Artificial Intelligence}
  \streetaddress{Trippstadter Str. 122}
  \city{Kaiserslautern}
%   \state{Germany}
  \country{Germany}
  \postcode{67663}
}

\renewcommand{\shortauthors}{Mengxi and Sizhen, et al.}

\begin{abstract}

This work described a novel non-contact, wearable, real-time eye blink detection solution based on capacitive sensing technology. A low-cost and low-power consumption capacitive sensing prototype was developed and deployed on a pair of standard glasses with a copper electrode attached to the glass frame. The eye blink action will cause the capacitance variation between the electrode and the eyelid. Thus by monitoring the capacitance variation caused oscillating frequency shift signal, the eye blink can be abstracted by a simple comparison of the raw frequency signal with a customized threshold. The feasibility and robustness of the proposed solution were demonstrated in five scenarios performed by eight volunteers with an average precision of 92\% and recall of 94\%. 

\end{abstract}




\maketitle

\section{Introduction}

Eye blink rate has been demonstrated to be an efficient indicator for fatigue and mental load\cite{martins2015eye} as well as other cognitive capacities\cite{jongkees2016spontaneous}. An example is the computer vision syndrome, which is a result of long-term screen stare and can be avoided by detecting the eye fatigue\cite{divjak2009eye} and supplying timely alerts. Besides that, eye blink has also been a hot topic in the human-computer interaction field as a novel input method, especially for patients with motor impairment, offering them a chance to interact with the digital world. Recent development in virtual reality and augmented reality also shows a demand for robust eye blink detection\cite{kumar2016electrooculogram}.

Existing eye blink detection approaches are mainly grouped into vision-based and sensor-based solutions, as Table \ref{related_work} lists. The vision solutions capture the landmarks of the face and abstract eye blink features with a deep convolutional model. Such a system provides high accuracy of eye blink detection and real-time inferences benefitting from the development of hardware resources and the feature abstracting deep models\cite{soukupova2016eye, panning2011color}. However, such systems are constrained in practical scenarios. Firstly, environmental conditions like light could cause a sharp drop in accuracy. Secondly, there are a lot of cases in everyday life where a camera does not exist when eye blink detection is needed. Thirdly, the vision solution is not a privacy-respect approach, thus, resulting in a low acceptance rate among users. On the other side, the sensor-based solutions do not face privacy issues since it captures only the eye-blink messages from the user. Jialin et al. \cite{liu2021blinklistener} proposed "BlinkListener" to sense the subtle eye blink motion using acoustic signals in a contact-free manner and demonstrated the robustness on the Bela platform as well as a smartphone with a median detection accuracy of 95\%. Elle et al. \cite{luo2020eslucent} introduced the Eslucent, an eyelid interface for detecting eye blinking, which is built upon the fact that the eyelid moves during the blink. By deploying an on-skin electrode on the eyelid, the capacitance value formed by the electrode and the eyelid will variate. The experiments show that the Eslucent achieves an average precision of 82.6\% and a recall of 70.4\%. Some other sensor-based approaches were also proposed for eye blink detection, like the doppler sensor\cite{kim2014detection}, EEG signals\cite{ko2020eyeblink}, mmWaves\cite{shu2022improved}, etc. However, the proposed sensor-based solutions lack either practicability, meaning that the long-term and in-the-wild eye blink detection is not possible, or robustness, meaning that the dynamic environmental conditions result in a limited number of usage scenarios. For example, the angular sensing range of the "BlinkListener" is limited. MmWave and Doppler sensors can not supply eye blink detection when people are out of the sensing field. EEG and the said "Eslucent" need special head/face preparation work before the detection.

\begin{table*}[t]
\centering
%\begin{threeparttable}
\footnotesize
\caption{Eye blink detection solutions}
\label{related_work}
\begin{tabular}{ |p{1.3cm}| p{1.3cm}| p{1.5cm}| p{1.5cm}| p{2.0cm}| p{2.0cm}| p{2.0cm}|}
%\toprule
\hline
authors/ year & Sensor  & Sensor Position & Performance & Algorithms & Advantages & limitation\\
%\midrule
\hline
Al-gawwam et al.\cite{al2018robust} 2018 & Camera & facing face & 96.65\% Precision &  Automatic facial landmarks detector, Savitzky–Golay Filter & High Precision & High computing load, Light Condition\\
\hline

Ryan et al.\cite{ryan2021real} 2021 & Event Camera & facing face & around 90\% Precision, subject dependent &  RNN & high temporal resolution & noise from background\\
\hline

De Lima et al.\cite{de2022efficient} 2021 & Web Camera & facing face & 92.63\% F-Score &  CNN + SVM & High F-score & High computing load, Light Condition\\
\hline
Jordan et al.\cite{jordan2020deep} 2020 & Camera in smart connected glass & facing face & 90.5\% Accuracy &  CNN & Edge Implementation & Light Condition\\
\hline

Luo et al.\cite{luo2020eslucent} 2020 & Capacitive sensor & eyelid & 82\% Precision &  Rule-based method & low power, wearable & low long-term detection capability \\
\hline

Maleki et al.\cite{maleki2018non} 2020 & Doppler sensor & in front of eye & not provided &  PCA & Robustness & low long-term detection capability \\
\hline

Cardillo et al.\cite{cardillo2021head} 2021 & mmWave sensor & in front of body & not provided &  interferometric algorithm & differiate between body movement and eye blink & very limited usage scenarios \\
\hline

Agarwal et al.\cite{agarwal2019blink} 2019 & EEG & head mounted EEG system & 93.4\% Precision &  automated unsupervised algorithm & high accuracy & EEG signal dependent \\
\hline

%\bottomrule
\hline
\end{tabular}
%\end{threeparttable}
\end{table*}

To address the existed sensing limitations, we proposed in this work a robust eye detection solution based on capacitive sensing technique. The solution was demonstrated with a series of experiments considering both static and dynamic body states. With the capacitive sensing unit deployed on a pair of standard glass, we achieved 92\% detection precision and 94\% recall. 

Overall, we have the following two contributions from this work:
\begin{enumerate}
\item A general concept of using capacitive sensing for eye blink detection in a non-contact way. 
\item A wearable and low-power eye blink detection system with demonstrated robustness and accuracy. Since the detection algorithm is simply implemented by comparison with a predefined threshold, the system supports real-time eye blink detection for long-term and in-the-wild usage scenarios.
\end{enumerate}

\section{Approach}
\subsection{General Principle}

\begin{figure}[hbt]
\includegraphics[width=0.45\linewidth, height = 3.0cm ]{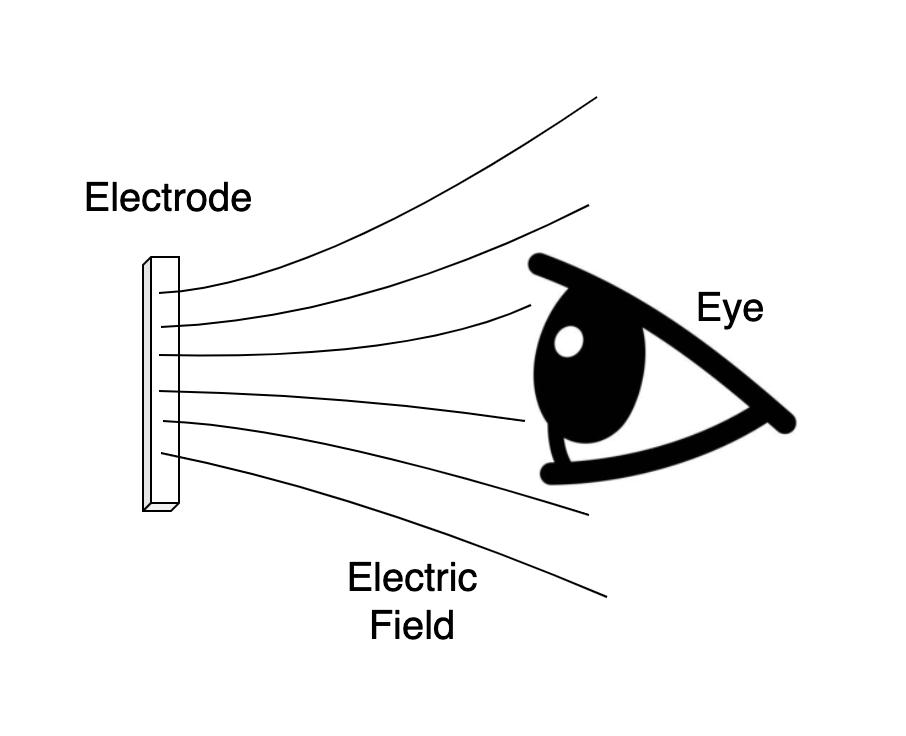}
\caption{Capacitance between Electrode and Eye}
\label{fig: capsensor}
\end{figure}

Capacitive-based proximity sensing relies on the electric field variations between/from electrodes caused by the environmental intrusion. Depending on the electrode number(single or pair) and the source signal(varying or static), three capacitive proximity sensing modes were widely explored: transmit mode\cite{ye2018capacitive}, shunt mode\cite{porins2019capacitive}, and load mode\cite{grosse2013opencapsense}. By measuring the capacitance variation, the proximity of the surrounding objects could be perceived. In this work, we designed a capacitive sensing unit that consists of a single electrode working in the loading mode to detect eye blinking. The design could be easily integrated into the glass frame. As Figure \ref{fig: capsensor} depicts, an electrode containing certain charges forms an electric field in the near surroundings. The blink action, namely the movement of the upper eyelid, will change the distance between the electrode and upper eyelid, disturbing the field distribution. Thus, the eye blink could be detected by monitoring the capacitance variation. 
Since the human body is a pure conductor, considering the entire body or body part as the conductive intrusion in the capacitive proximity sensing system, a wide range of movement types caused by the entire body\cite{bian2019wrist} or body part\cite{bian2021capacitive} could be sensed, which enables a new branch of wearable motion sensing modality besides the dominant inertial sensor unit\cite{bian2022using}.

\subsection{Hardware Implementation}

Since body-area capacitance value or its variation is not as easy as other parameters like voltage signal to be measured directly, researchers have explored indirect measurement solutions. Two kinds of solutions are commonly used: the charge-based and frequency-based methods. The former one monitors the body-area capacitance by measuring the charge variation caused by voltage signal\cite{bian2019passive, braun2011capfloor}, which usually needs a pre-processing step like amplification with orders of magnitude or high-resolution sampling for the capture of subtle raw signal;  the latter one records the body-area capacitance by counting the capacitance variation caused frequency shift signal in an oscillating circuit, which could be implemented directly by a counter\cite{grosse2012enhancing, bian2021systematic}. Considering the cost and complexity, we chose the latter solution for eye-blink detection. The oscillator is an LC tank, and the frequency is measured and digitized by the chip FDC2214, which also drives the LC tank to oscillate and supplies four channels for capacitance variation monitoring. Currently, only one channel is utilized in this work, and our further exploration will use multiple channels for extended eye activity detection. A capacitance change in the LC tank will be observed as a shift in the resonant frequency. In the following description, we use frequency data directly for blink action abstraction instead of backward computing the capacitance variation value.

\begin{figure}[hbt]
\includegraphics[width=0.7\linewidth, height = 6cm ]{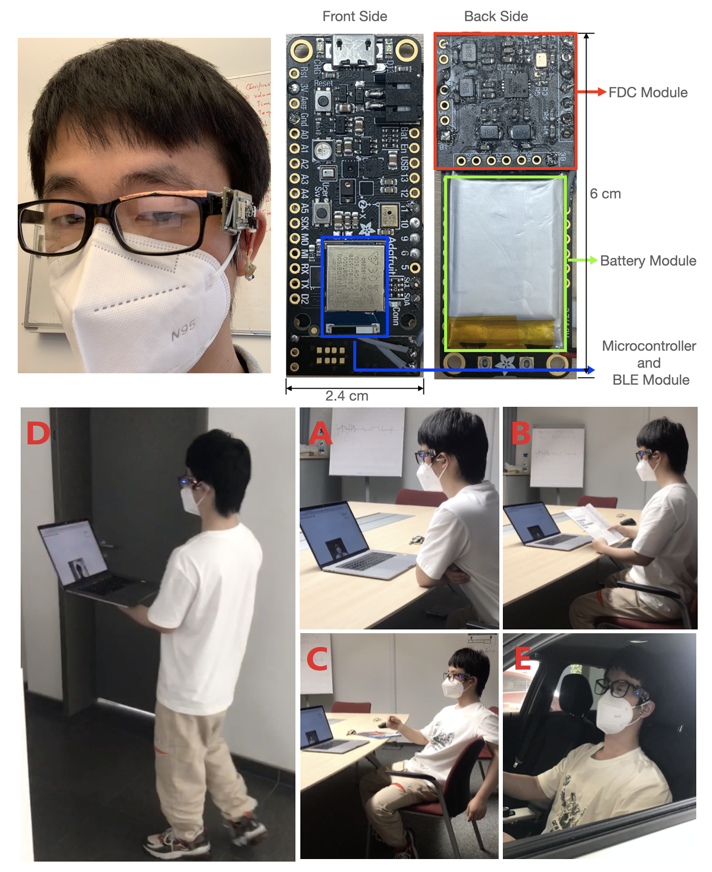}
\caption{Top: Hardware prototype and its deployment on the glass. Bottom: Five eye blink detection scenarios, A: intentional blink; B: Involuntary blink while reading a book; C: Involuntary blink while talking(replaced by reading a news with voice in some sessions); D: Involuntary blink while walking around; E: Involuntary blink while sitting in the car cockpit simulating the driving activity}
\label{fig: combined}
\end{figure}

Besides the sensing unit, the hardware platform includes an nRF52840 micro-controller for real-time eye blink abstraction, which also supports Bluetooth transmission for offline data presentation. A compact lithium battery with a 300 mAh capacity is used to power the system, supporting about $20$ hours of continuous in-the-wild eye blink detection. The electrode is simply a piece of copper tape attached to the upper frame of a pair of standard glasses. The whole hardware prototype wights only around $18$ g that users barely feel when wearing. Figure \ref{fig: combined}(top) shows the hardware platform and its deployment on the glass.

\section{Evaluation}
\label{sec:evaluation}
To evaluate the feasibility and robustness of the proposed eye blink detection approach and hardware prototype, five experiments were performed regarding different daily activities, wherein one intentional and four involuntary blink scenarios were recorded, as Figure \ref{fig: combined} depicts. Since not every volunteer owns a driver's license, the scenario in the car cockpit was recorded in a parking lot. In addition, during the talking scenario, some volunteers simulated talking by reading a piece of random news with the voice in case there was no talking topic between the volunteer and the operator.

Eight volunteers (five males and three females) aged 24 to 32 participated in the planned scenarios. Each scenario takes around eight minutes with a data sampling rate of 60Hz. A web application written with Javascript was designed to capture, plot, store, and label the data transmitted from Bluetooth. Meanwhile, the web application activates the PC camera to record the face of the volunteer during the experiment, aiming to abstract the eye blink ground truth.

\begin{figure}[hbt]
\includegraphics[width=\linewidth, height = 4cm ]{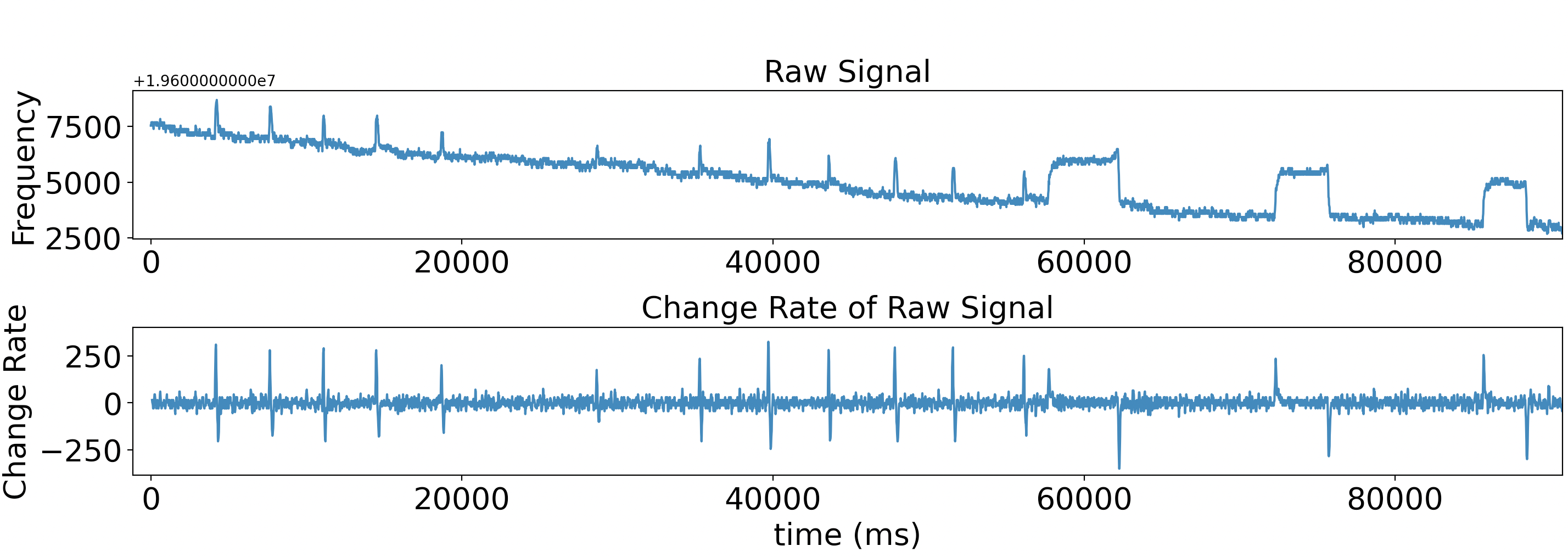}
\caption{Raw signal and signal variation between two continuous raw data}
\label{fig: sensitive test}
\end{figure}

Study\cite{kwon2013high} shows that during a voluntary blink, the closing phase takes about 100 ms, and the opening phase takes about 200 ms until 97 percent recovery of opening. The capacitance variation occurs essentially during the closing and opening phases. The closed and opened phases of the eye will not change the capacitance since there is no distance variation between the electrode and the eyelid. Thus we used a one-second sliding window with a half-second overlap for real-time blink detection. A low-pass filter was first applied to the raw data to filter out the inherent noise. Then we calculated the data variation by simply subtracting two continuous raw data. This step removes the data drift caused by environmental electric field variation and outputs distinctive features of eye blinks. Figure \ref{fig: sensitive test} shows an example of the raw data stream and data variation stream. Each peak in the raw data stream represents a blink action. In the data variation stream, the positive peaks represent the closing phase of a blink action, and the negative peaks represent the opening phase. For blink detection, we set a predefined threshold for comparison to capture the positive impulse in the signal variation stream. The last three square waves in the raw signal stream indicate that the prototype could also perceive the duration of the eye-closing state, which will be described in detail in our future work.

\begin{enumerate}[font=\bfseries,label=(\Alph*)]\setcounter{enumi}{0}
\item \textbf{Intentional Eye Blink}
\end{enumerate}

\begin{figure}[hbt]
\includegraphics[width=\linewidth, height = 4cm ]{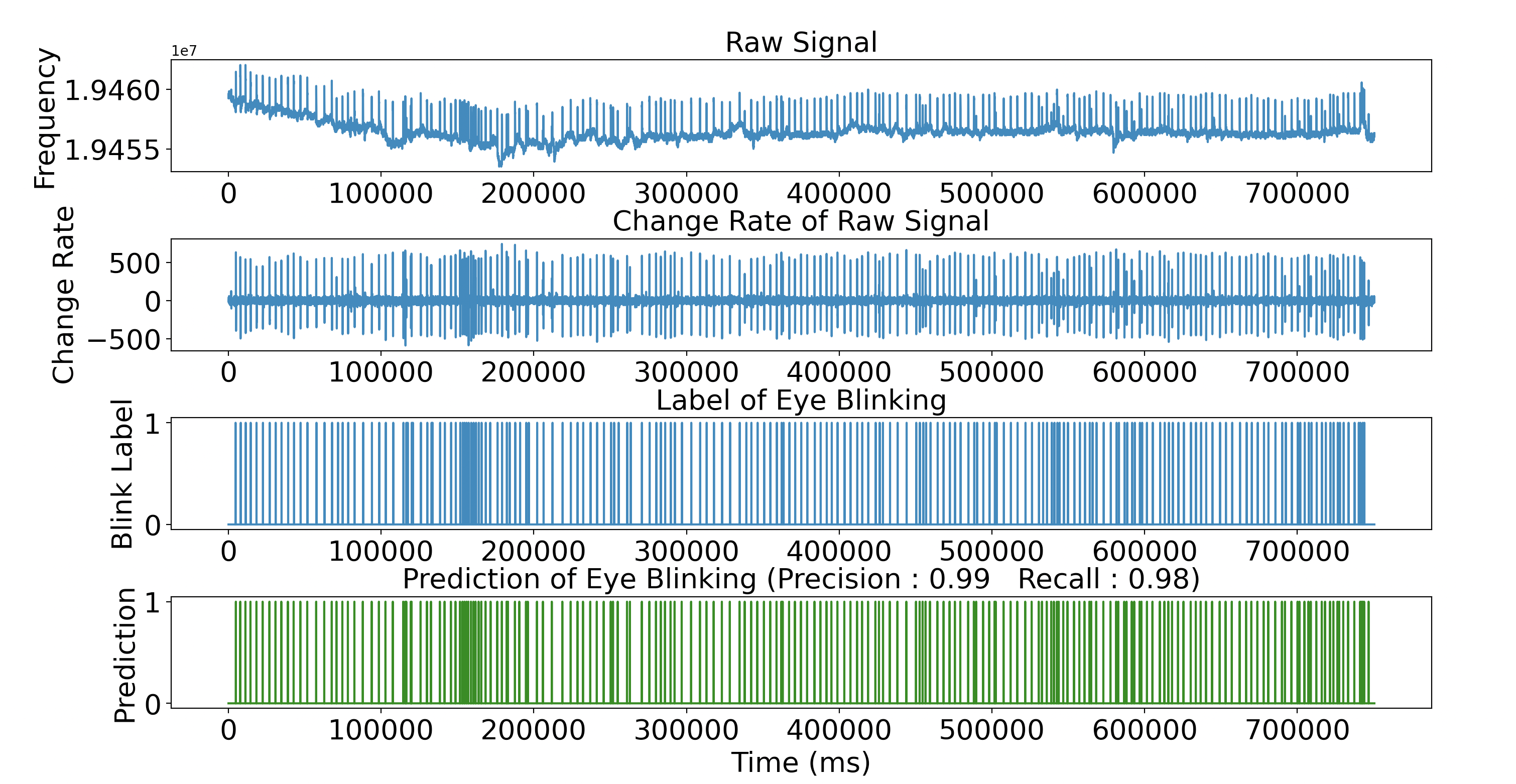}
\caption{Raw signal and signal variation of the intentional blink scenario together with the ground truth and the detected blink(green)}
\label{fig: Intentional blinking}
\end{figure}

We first carried out the intentional eye blink experiment aiming to supply a baseline for the followed involuntary experiments. Volunteers were asked to sit in front of the laptop wearing the prepared glasses and blinking eyes intentionally for around eight minutes. The laptop was used to receive and plot the raw data stream from the hardware prototype deployed on the glasses, meanwhile recording a video from which the ground truth of eye blink can be labeled. The blinking frequency is about ten times per minute. Figure \ref{fig: Intentional blinking} shows the raw signal and signal variation of the intentional blink together with the ground truth and the detected blink(green) from one of the volunteers. The drift in the raw signal relates to the capacitance between the electrode and the environment. For example, a head movement will result in a noticeable drift variation. However, since eye blink happens fast and the capacitance change rate of eye blink signal is large enough, the impulses caused by eye blink are still visibly distinctive. By simply comparing the signal variation with a predefined threshold, we detected the eye blink with a precision of 0.99 and a recall of 0.98 from the volunteer. The averaged intentional blink detection gives 0.93 precision and 0.94 recall.

\begin{enumerate}[font=\bfseries,label=(\Alph*)]\setcounter{enumi}{1}
\item \textbf{Involuntary Eye Blink}
\end{enumerate}

\begin{figure}[hbt]
\includegraphics[width=\linewidth, height = 4cm ]{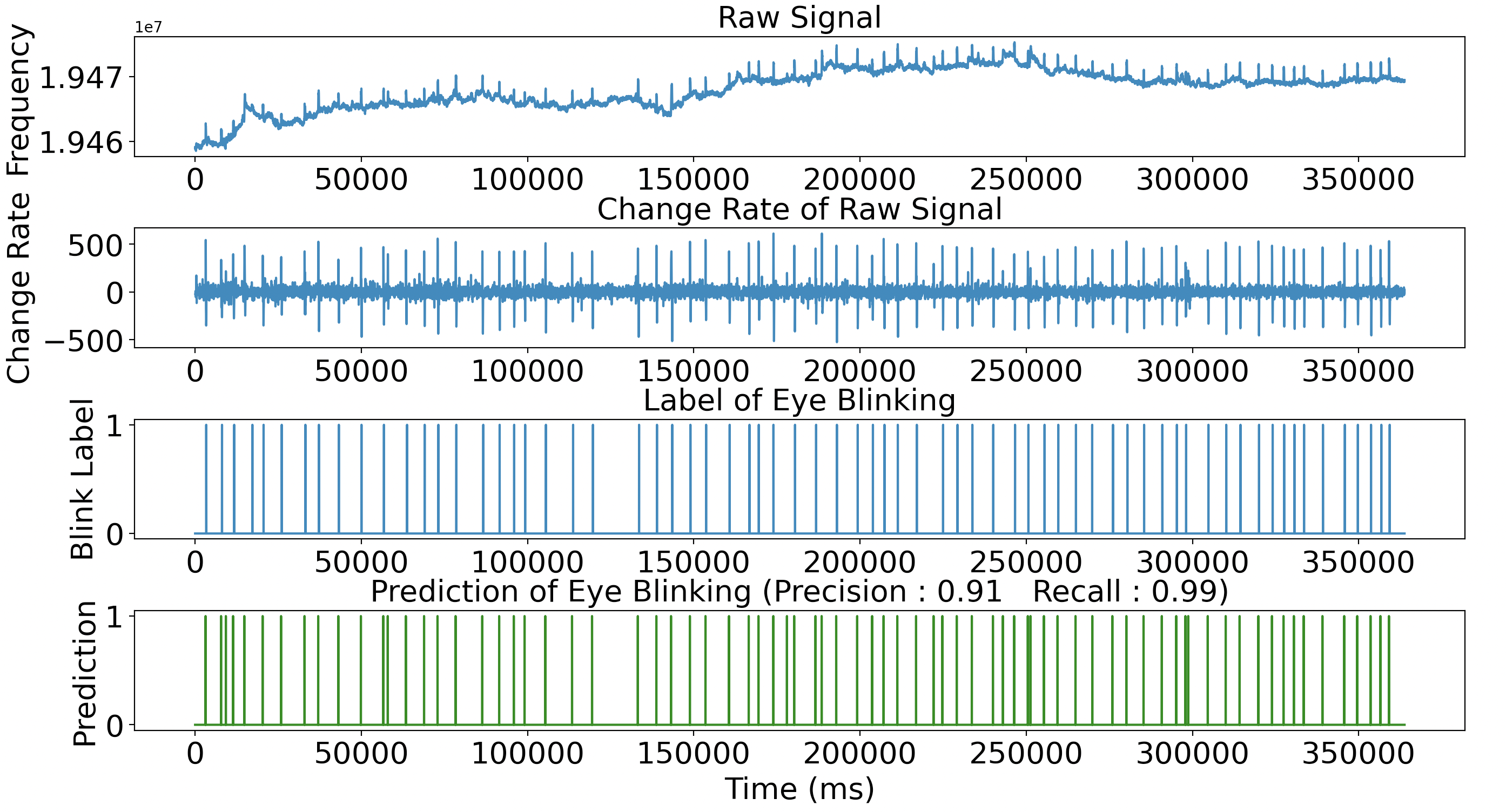}
\caption{Raw signal and signal variation of the walking scenario together with the ground truth and the detected blink(green)}
\label{fig: walking}
\end{figure}

To demonstrate the feasibility of the proposed prototype in practical life scenarios like reading and talking, all volunteers performed four everyday activities wherein both static and dynamic body states were considered. A simulated driving scenario was carried out in a parking lot, considering that involuntary eye blink monitoring has been proved efficient for fatigue indication\cite{wang2006driver} during driving. In the talking scenario, volunteers wore glasses and chatted with the operator or read out a piece of random news. The reading activity was recorded during reading a paper or book or operating the smartphone in silence. In the Walking scenario, all volunteers wore glasses and walked inside our office building. They walked in a random direction to check the robustness of our prototype in different environmental surroundings. 
Figure \ref{fig: walking} shows the signals, labels, and detected blinks during the walking of one volunteer. As can be seen, the eye blink action still gives evident impulse in the signal variation stream, and an impressive precision(0.91) and recall(0.99) are achieved using the proposed positive peak detection approach. A similar achievement is also presented in the other three explored daily activities.

Table \ref{tab:results} lists the result of eye blink detection with our prototype among the volunteers, wherein V1, V2, and V6 give impressive results with an averaged precision and recall of over 0.95. V6 shows an average recall of 0.98, meaning that nearly all blinks were detected, especially in the walking and driving activity, with a recall of 0.99. V5 delivers an averaged precision of 0.86, meaning that a number of false positives were detected, mainly during the involuntary eye blink scenarios (0.91 precision in the intentional blink). The averaged precision and recall from all the volunteers in all activities are 0.92 and 0.94, respectively.    

\begin{table}[hbt]
\centering
\footnotesize
\caption{Eye blink detection performance(precision/recall, \%)}
\label{tab:results}
\begin{tabular}{|p{1.0cm}|p{0.45cm}|p{0.45cm}|p{0.45cm}|p{0.4cm}|p{0.4cm}|p{0.4cm}|p{0.4cm}|p{0.4cm}|p{0.9cm}|}
\hline
Volunteer & V1 & V2 & V3 & V4 & V5 & V6 & V7 & V8 & Averaged\\
%\bottomrule
\hline
Intentional Blink & 99/99 & 100/98  & 84/93 & 89/89 & 91/94 & 99/98 & 94/94 & 88/90 & 93/94\\

\hline

Reading              & 94/98 & 96/97  & 95/100 & 94/94 & 93/95 & 89/95 & 99/97 & 86/86 & 93/95\\
\hline

Talking              & 100/97 & 95/97  & 85/98 & 88/93 & 85/96 & 97/98 & 92/89 & 89/88 & 91/95\\

\hline

Walking              & 99/97 & 90/92  & 84/85 & 95/94 & 81/86 & 91/99 & 95/98 & 94/98 &  91/94\\

\hline

Driving              & 89/90 & 99/94  & 88/93 & 95/99 & 79/92 & 99/99 & 83/83 & 88/96 & 90/93\\
\hline
Averaged             & 96/96 & 96/96  & 87/94 & 92/94 & 86/93 & 95/98 & 93/92 & 89/92 & 92/94\\

\hline

\end{tabular}
\end{table}

\section{Limitation}
Although the prototype shows competitive blink detection performance, we observed a few limitations of the proposed approach for eye blink detection during the experiments. First, two pieces of glasses of different sizes composed of plastic frames were prepared for volunteers who do not wear a glass, and the blink detection performance degrades if the glass doesn't match the volunteer's head form. This is reasonable since the proposed approach is based on the approaching of the upper eyelid to the capacitive electrode. A closer distance between the eyelid and electrode gives better sensitivity. Thus the volunteers who attended the experiment were selected when their head form matched the glass frames well. Second, since each volunteer's blink signal variation scale is different, the threshold used for peak detection among the volunteers is not a constant value, which seems to be affected by the eye form and the distance between eye and electrode. A proper threshold should be adaptive, varying alongside the signal stream, which is achievable by analyzing the signal variation in real-time. Third, intensive body actions like running cause the glass vibration relative to the head, which results in a more significant signal than the blink-caused signal. This work did not explore blink detection in such scenarios with strenuous intensity, but it will be one direction for future work.

\section{Conclusion and Future Work}

This work demonstrates a non-contact, real-time eye blink detection solution based on the capacitive sensing modality. By deploying a sensitive capacitive-based proximity sensor with the electrode on the glass frame, an impulse-like signal can be observed during each blink since the upper lid movement is approaching or departing the electrode. The blink action is captured by monitoring the oscillating frequency shift caused by the capacitance variation. Eight volunteers performed intentional and involuntary blinks during a few daily activities and showed an averaged blink detection performance with 0.92 precision and 0.94 recall. To explore the additional capability of the proposed approach in the future, multiple electrodes supported by multiple sensing channels will be deployed around the glass frame, aiming to sense the blink of each eye and the duration of opened/closed eye state. Furthermore, a sensing fusion approach will also be tried to sense the blink during intensive activities like running.   

\section{Acknowledgement}
This work has been supported by the BMBF (German Federal Ministry of Education and Research) in the project Eghi (number 16SV8527).

\bibliographystyle{ACM-Reference-Format}
\bibliography{sample-base}

\end{document}